# On Enforcing Database Constraints Using MS VBA Event-Driven Procedures and SQL Server

**Diana Christina Mancas**

**Abstract**

The goal of this paper is to provide a rigorous methodology for enforcing database constraints using MS VBA event-driven procedures for software applications built on top of MS SQL Server databases, in the framework of the software engineering Database Constraint-Driven Design and Development. The original contribution is a pseudo-code algorithm for assisting developers in this process. We exemplify the results of using it with some VBA code examples taken from a genealogy database software application designed and developed using this methodology. Consequently, it enforces all the database constraints -be they relational or not- which govern this sub-universe of discourse, thus guaranteeing the highest possible quality of its managed data. We also show that the proposed algorithm works fine not only for other relational database management systems, but even with NoSQL platform backends, and that the modifications needed to adapt it to other similar SQL embedding frontend platforms are minimal.

## 1 Introduction

In Mancas et al. [1] a Database Constraint-Driven Design and Development (DCDDD) methodology was introduced as a software engineering paradigm for guaranteeing the highest possible quality of the data managed by software applications. This methodology was very successfully used for the design and development of a software database (db) application managing genograms [2], with which the DCDDD approach was exemplified in [1].

DCDDD has the following 6 steps, in this order: sub-universe analysis; translation of the resulting Entity-Relationship data model [3] into a (Elementary) Mathematical Data Model ((E)MDM) [4] scheme; (E)MDM scheme validation and enhancement; corresponding Relational Data Model (RDM) [3] db generation; corresponding db software application generation driven by the non-relational db constraints; ergonomic polishing of the generated application Graphic User Interface (GUI).

The core last but one step has the following 3 sub-steps, in this order: for all constraints, detect all use cases in which they might be violated; establish the corresponding event-driven procedures needed to be coded; generate needed code for enforcing all db constraints that cannot be enforced by the host Database Management System (DBMS).

The first four steps, as well as the first sub-step of the fifth one are extensively discussed and illustrated in [1] with examples from [2]. The latter two sub-steps of the core step are only generally approached in [1], as they are heavily dependent on the technological platforms used. This paper presents them in detail for the MS VBA and SQL Server platforms, with examples from [2], as well.

The *GENEALOGIES* [2] db software application manages genograms, i.e., genealogical trees extended with geographical, architectural, and touristic data. In the first DCDDD three steps, we identified 210 business rules that are governing this sub-universe [1, 2] and formalized them as db constraints. 172 of them are of the six RDM types: domain (range); not null; uniqueness (key); referential integrity (foreign key); tuple (check); and default values. The remaining 38 constraints are non-relational, but of the four (E)MDM categories: set; dyadic relationship; function; object.

Unfortunately, 19 relational constraints cannot be enforced by the MS SQL Server: 15 of them are keys that contain at least one column that is not null, and which might have more than one null value, and 4 are on VARBINARY objects used to store, via OLEDB embedding, high-resolution pictures of cities and monuments. Out of the latter 4, only 2, those of the not-null type, were enforced through VBA: those of the uniqueness type (i.e., no city or monument picture should be stored more than once) cannot be enforced through VBA either, as not even its Variant data type is capable of handling such huge memory objects.

Consequently, the VBA frontend *GENEALOGIES* application enforces 55 constraints, out of which 17 are relational and 38 are non-relational. VBA, the MS *Office Suite* programming language, was chosen especially for its rich panoply of db related event-driven procedure types, because it embeds SQL, but also for its simplicity and robustness.

## 1.1 Related work

The VBA programming platform is fully documented in [5]. There are thousands of papers on event-driven programming (e.g., [6 – 10]), a paradigm that arose from the finite state machines field.

Unfortunately, there are much fewer papers and books on non-relational database constraint enforcement.

Non-relational database constraints may be enforced in two ways: either using triggers of the DBMS' extended SQL (e.g., IBM's SQL PL, Oracle's PL/SQL, MS' T–SQL, etc.) or using event-driven procedures of a high-level programming language embedding SQL (e.g., MS VBA, C#, ASP, etc.).

For example, [11, 12] advocate the first one of these approaches, while [13, 14] advocate the second one. We prefer the second approach, for several reasons, namely:

1. Why passing to the backend DBMS unplausible values, when the frontend can much faster detect and reject them?
2. Generally, triggers are much slower than event-driven procedures.
3. Extended SQLs differ significantly between them (as there is no agreed extension standard), which makes triggers very hard to port from one DBMS to another. On the contrary, thanks to the *de facto* standards of ADOX and ODBC, any db software application needs only an almost insignificant programming effort to adjust its frontend to virtually any other relational DBMS.

## 1.2 Paper outline

The next section is the core of our original contribution, as it presents our proposed pseudo-code algorithm aimed to assist the design and development in VBA of the event-driven procedures needed to enforce both non-relational and relational constraints that cannot be enforced by the SQL Server (where ' followed by a space starts comments).

Section 3 provides corresponding code examples from [2], within the limits of the maximum allowed paper length: two apparently simple constraints were chosen for their coding complexity. The paper ends with a concluding and further work section, as well as references.

## 2 The algorithm for db constraint enforcement in VBA

*Algorithm VBAConstraintEnforcement*

*Input*: a constraint *c* and the set of forms $\{F_1, \ldots, F_n\}$, $n > 0$, containing at least one control whose data source is a db table column involved in *c*
*Output*: the set of forms $\{F_1, \ldots, F_n\}$ having their corresponding classes augmented with the VBA code needed for enforcing *c* as well

*Strategy*:
*If c* might be violated by time passing *Then FL*(*c*)
*For i* = 1 *To n*
  *If c* involves only column *f Then*
    *If f* is not a foreign key *Then ctrlBU*($c$, $f$, $F_i$) *Else FC*($c$, $f$, $F_i$) *End If*
  *ElseIf c* involves only columns *f* and *g* and *f* is the link column of sub-form $F_i$ *Then*
    *ctrlBU*($c$, $g$, $F_i$) *Else FBU*($c$, $F_i$)
  *End If*
  *FAU*($c$, $F_i$)
  *FADC*($c$, $F_i$)
*Next i*
*End Algorithm VBAConstraintEnforcement*

Fig. 1. The proposed pseudo-code algorithm for assisting db constraint enforcement in VBA

*FL*(*c* AS Constraint)

‘ Use the *Form_Load* event procedure of the application’s *Menu* form to check and correct the db
‘ instance accordingly, if needed.

*If* the application’s *Menu* form does not have an associated *Form_Load* event procedure
*Then* add it to its class;
add in *Form_Load*() procedure of class *Menu* the code to check whether *c* is violated;
*If* this is the case *Then*
  adjust corresponding application parameters so that *c* is satisfied;
  display corresponding warning message;
*End If*
*End Sub FL*

Fig. 2. The method *FL* called by Algorithm *VBAConstraintEnforcement* from Fig. 1

*ctrlBU*(*c* AS Constraint, *f* As Control, $F_i$ As Form)

‘ Use the *f_BeforeUpdate* event procedure to reject unplausible values.

*If* the application’s $F_i$ form does not have an associated *f_BeforeUpdate* event procedure
*Then* add it to its class;
add in *f_BeforeUpdate*(*Cancel* As Integer) procedure of class $F_i$ the following code:

```
If Not Cancel And f <> f.OldValue Then
   If f violates c Then
      Cancel = True
      Beep
      MsgBox “Corresponding error message”, vbCritical, “Request rejected…”
      f.Undo
   End If
End If
```

*End Sub ctrlBU*

Fig. 3. The method *ctrlBU* called by Algorithm *VBAConstraintEnforcement* from Fig. 1

*FC*(*c* AS Constraint, *f* As Control, $F_i$ As Form)

‘ Use the *Form_Current* event procedure to eliminate unplausible values from combo-box *f*.

*If* the application’s $F_i$ form does not have an associated *Form_Current* event procedure
*Then* add it to its class;
inject in the WHERE clause of the SQL SELECT statement that is the row source of *f* the code needed to eliminate from the *f*’s combo-box instance, for the current data row, all values that would violate *c* and save the resulting SELECT statement in a local string variable *s*;
add in *Form_Current*() procedure of class $F_i$ the following code:

```
f.RowSource = s
f.Requery
```

*End Sub FC*

Fig. 4. The method *FC* called by Algorithm *VBAConstraintEnforcement* from Fig. 1

*FBU*(*c* AS Constraint, $F_i$ As Form)

‘ Use the *Form_BeforeUpdate* event procedure to reject unplausible values.

*If* the application’s $F_i$ form does not have an associated *Form_BeforeUpdate* event procedure
*Then* add it to its class;
add in *Form_BeforeUpdate*(*Cancel* As Integer) procedure of class $F_i$ the following code (where $f_1$, …, $f_m$ are the *m* columns involved in *c*, $m > 0$):

```
If Not Cancel And (f1 <> f1.OldValue Or … Or fm <> fm.OldValue) Then
   If f1, …, fm violate c Then
      Cancel = True
      Beep
      MsgBox “Corresponding error message”, vbCritical, “Request rejected…”
      f1.SetFocus
   End If
End If
```

*End Sub FBU*

Fig. 5. The method *FBU* called by Algorithm *VBAConstraintEnforcement* from Fig. 1

*FAU*(*c* AS Constraint, $F_i$ As Form)

‘ Use the *Form_AfterUpdate* event procedure to synchronize data.

*If* the application’s $F_i$ form does not have an associated *Form_AfterUpdate* event procedure
*Then* add it to its class;
add in *Form_AfterUpdate*() procedure of class $F_i$ the code needed to enforce *c*;
*End Sub FAU*

Fig. 6. The method *FAU* called by Algorithm *VBAConstraintEnforcement* from Fig. 1

*FADC*(*c* AS Constraint, $F_i$ As Form)

‘ Use the *Form_AfterDelConfirm* event procedure to synchronize data.

*If* the application’s $F_i$ form does not have an associated *Form_AfterDelConfirm* event procedure
*Then* add it to its class;
add in *Form_AfterDelConfirm*(*Status* As Integer) procedure of class $F_i$ the following code:

```
If Status = acDeleteOK Then
   ‘ add here code needed to enforce c
End If
```

*End Sub FADC*

Fig. 7. The method *FADC* called by Algorithm *VBAConstraintEnforcement* from Fig. 1

# 3 VBA code examples of enforcing non-relational constraints

The structural E-R diagram of the *GENEALOGIES* db [2] is the following:

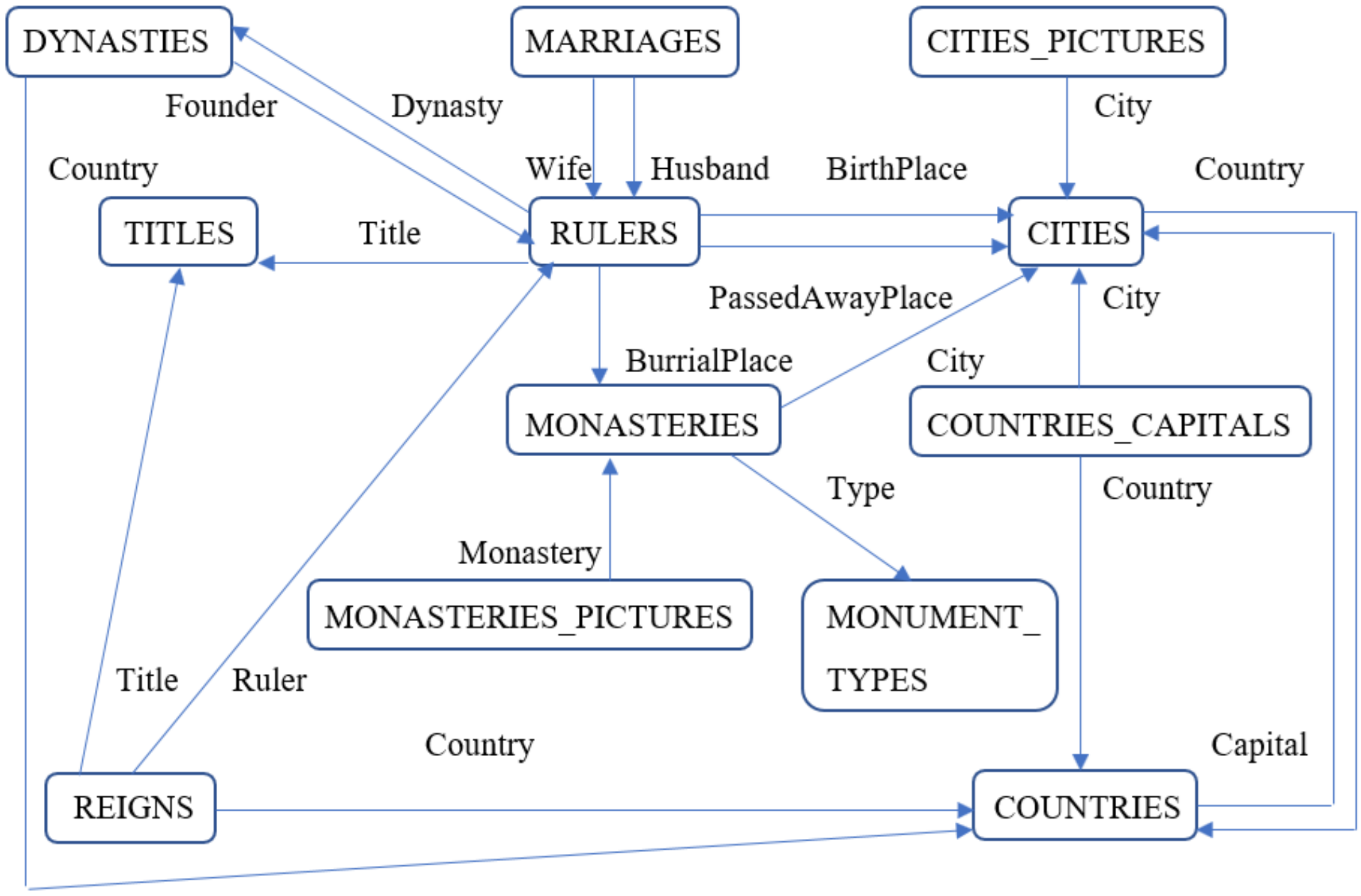

## 3.1 Constraint $C_4$ (nobody may live less than 0 or more than *maxLifeYears*)

This constraint might be violated either by the time that passes (as for living persons their passed-away date is null) or by changing persons' birth or/and passed away dates [1, 2]. In theory, it is relational, but SQL Server may not enforce it as all 6 involved columns may store any number of null values. The corresponding VBA code for enforcing it is presented in Fig. 8, 9 and 10.

### 3.1.1 Code added to the *Form_Load*() method of class *Menu*

```
‘ enforce constraint C4
Dim maxAge As Variant
Dim Name As String
maxAge = DMax("Age", "Ages")          ‘ uses view Ages - see Fig. 9
If Not IsNull(maxAge) Then
  If maxAge > maxLifeYears Then       ‘ maxLifeYears is one of the app.’s parameters, stored in db table PARAMS
    Beep
    Name = DLookup("Name", "Ages", "Age =" & maxAge)
    MsgBox "At least " & Name & "'s age is " & maxAge & " years old, which is greater than " & maxLifeYears & _
        ", the current value of the parameter maxLifeYears: consequently, maxLifeYears has been automatically" & _
        " increased to & CInt(maxAge) + 1 & ", such as to satisfy constraint C4!", vbInformation, "Constraint C4 " & _
        "was violated by the passing time!"
    maxLifeYears = CInt(maxAge) + 1
    DoCmd.RunSQL "UPDATE PARAMS SET maxLifeYears =" & maxLifeYears
  End If
End If
```

Fig. 8. The VBA code added to the *Form_Load* procedure of class *Menu* to enforce constraint $C_4$

This piece of VBA code calls the SQL view *Ages* (see Fig. 9), which computes the ages of all persons stored in the db (assigning -1 for those whose birth date is unknown).

This code is so complex because the *GENEALOGIES* application allows users to store data on persons born since the year 6,000 BC, with separate columns for their years, months, and days of both birth and death, as some of them might be only partially known (cases in which the application replaces months and days with guesses), as well as because VBA considers the minimum calendar date value as being January 1st 100 (also see Fig. 10).

**3.1.2 Code added to the *Form_BeforeUpdate*(*Cancel* As Integer) method of class *PERSONS***

The *Form_BeforeUpdate* procedure from Fig. 10 was chosen because constraint $C_4$ concerns 6 columns of this table, namely *BirthYear*, *BirthMonth*, *BirthDay*, *PAYear*, *PAMonth*, and *PADay*. In fact, this code is also added to the *Form_BeforeUpdate* procedures of the classes of the *PERSONS* sub-forms *SIBLINGS*, *SIBLINGSofMOTHER, CHILDREN*, and *CHILDRENofMOTHER*, which are also based on the db table *RULERS* storing persons data.

```
SELECT Name, IIf([PassedYear]<0,Cdbl(Round((CDate(IIf(IsNull([BirthMonth]),"7",CStr([BirthMonth])) & "-" &
       IIf(IsNull([BirthDay]),"1",CStr([BirthDay])) & "-" & CStr(IIf([PassedYear]<=-100,-[BirthYear],100-[BirthYear])))-
       CDate(IIf(IsNull([PassedMonth]),"6",CStr([PassedMonth])) & "-" & IIf(IsNull([PassedDay]),"28",CStr([PassedDay])) &
       "-" & CStr(IIf(PassedYear<=-100,-PassedYear,100-PassedYear))))/365,2)),IIf(BirthYear<0,Cdbl(Round(((CDate(IIf(
       IsNull(BirthMonth), "7",CStr(BirthMonth)) & "-" & IIf(IsNull(BirthDay),"1",CStr(BirthDay)) & "-" & CStr(IIf(BirthYear<=-
       100,-[BirthYear],100-[BirthYear])))-#1/1/100#)+(CDate(IIf(IsNull([PassedMonth]),"6",CStr([PassedMonth])) & "-" &
       IIf(IsNull(PassedDay),"28",CStr(PassedDay)) & "-" & CStr(IIf([BirthYear]<=-100,[PassedYear],100+[PassedYear])))-
       #1/1/100#))/365,2)),Cdbl(Round((CDate(IIf(IsNull([PassedMonth]),"6",CStr(PassedMonth)) & "-" & IIf(IsNull(
       PassedDay),"28",CStr(PassedDay)) & "-" & CStr(IIf[BirthYear<100,PassedYear+100,PassedYear)))-CDate(IIf(
       IsNull(BirthMonth),"7",CStr(BirthMonth)) & "-" & IIf(IsNull(BirthDay),"1",CStr(BirthDay)) & "-" & CStr(IIf(BirthYear<
       100,BirthYear+100,BirthYear))))/365,2)))) AS Age, BirthYear, BirthMonth, BirthDay, PassedYear, PassedMonth,
       PassedDay, [#R]
From RULERS
WHERE NOT BirthYear Is Null AND NOT PassedYear Is Null
UNION
SELECT Name, -1, BirthYear, BirthMonth, BirthDay, PassedYear, PassedMonth, PassedDay, [#R] FROM RULERS
WHERE BirthYear Is Null
UNION
SELECT Name, Cdbl(Year(Date()) - BirthYear), BirthYear, BirthMonth, BirthDay, PassedYear, PassedMonth, PassedDay,
       [#R]
FROM RULERS
WHERE NOT BirthYear Is Null and PassedYear Is Null
ORDER BY Age DESC, Name, [#R];
```

Fig. 9. The SQL code of view *Ages* invoked by the VBA code from Fig. 8

```
' enforce constraint C4
Dim by, bm, bd, pay, pam, pad As Integer           ' birth and passed away years, months, and days
Dim d1, d2 As Date                                 ' adjusted min and max life dates
Dim d As Long                                      ' number of lived days for the current person
If Not Cancel And Not IsNull(BirthYear) And (BirthMonth <> BirthMonth.OldValue Or BirthDay <>
       BirthDay.OldValue Or PAYear <> PAYear.OldValue Or PAMonth <> PAMonth.OldValue Or PADay <>
       PADay.OldValue) Then
  by = BirthYear
  bm = IIf(IsNull(BirthMonth), 6, BirthMonth)      ‘ when not known birth month, assume it was June
  bd = IIf(IsNull(BirthDay), 1, BirthDay)          ‘ when not known birth day, assume it was 1st of bm
  pay = IIf(IsNull(PassedYear), Year(Date), PassedYear) ‘ current system year for persons alive
  pam = IIf(IsNull(PassedMonth), 7, PassedMonth)   ‘ when not known death month, assume it was July
  pad = IIf(IsNull(PassedDay), 28, PassedDay)      ‘ when not known death day, assume it was 28 of pam
  If pay < 0 Then
    If pay < -100 Then
      by = -by
      pay = -pay
    Else
      by = 100 - by
      pay = 100 - pay
    End If
```

Fig. 10. The VBA code added to the *Form_BeforeUpdate* procedure of class *PERSONS* to enforce constraint $C_4$

```
    d1 = CDate(pam & "-" & pad & "-" & pay)
    d2 = CDate(bm & "-" & bd & "-" & by)
  Else
    If by < 100 Then                          ‘ translate birth and death dates with 300 years ahead
      by = by + 300
      pay = pay + 300
    End If
    d1 = CDate(bm & "-" & bd & "-" & by)
    d2 = CDate(pam & "-" & pad & "-" & pay)
  End If
  d = DateDiff("d", d1, d2)
  If d < 0 Then
    Cancel = True
    Beep
    MsgBox "Nobody may live less than 0 days!",vbCritical, "Please correct birth or/and passed away dates..."
  ElseIf d / 365.2425 > maxLifeYears Then
    Cancel = True
    Beep
    MsgBox "Nobody may live more than " & maxLifeYears & " years!", vbCritical, _
           "Please correct birth or/and passed away dates..."
  End If
  If Cancel Then BirthYear.SetFocus
End If
```

Fig. 10 (Continued)

## 3.2 Constraint $C_{20}$ (For any marriage, both spouses must be simultaneously alive for at least one day.)

This constraint might be violated either by selecting as spouse of a marriage somebody who does not satisfy this condition or by modifying marriage and/or divorce dates or by modifying birth and/or death dates of a spouse [1, 2]. The corresponding VBA code is presented in Fig. 11, 12, 13, and 14.

### 3.2.1 Code added to the *Form_BeforeUpdate*(*Cancel* As Integer) method of class *PERSONS*

```
' enforce constraint C20
Dim q As String
Dim rs As ADODB.Recordset
' birth date > spouse passed away date?
If Not Cancel And Not NewRecord And Sex <> "N" And Not IsNull(BirthYear) Then
  If Not IsNull(DLookup("[#M]", "MARRIAGES", "Husband =" & [#R] & " Or Wife =" & [#R])) Then
    Set rs = New ADODB.Recordset          ' current person had at least one spouse
    If Sex = "M" Then                     ‘ compute the set of wives for the current male person
      q = "SELECT Name, PassedYear, PassedMonth, PassedDay FROM MARRIAGES INNER JOIN RULERS " _
              & "ON MARRIAGES.Wife = RULERS.[#R] WHERE Husband =" & [#R]
    Else                                  ‘ compute the set of husbands for the current female person
      q = "SELECT Name, PassedYear, PassedMonth, PassedDay FROM MARRIAGES INNER JOIN RULERS " _
              & "ON MARRIAGES.Husband = RULERS.[#R] WHERE Wife =" & [#R]
    End If
    rs.Open q, Application.CurrentProject.Connection, adOpenForwardOnly, adLockReadOnly
    While Not rs.EOF And Not Cancel       ‘ check passed away dates spouse by spouse
      If Not IsNull(rs!PassedYear) Then   ‘ null values may not violate this constraint
        If rs!PassedYear < BirthYear Then ‘ spouse dead before current person’s birth?
          Cancel = True
        ElseIf rs!PassedYear = BirthYear Then
          If Not IsNull(rs!PassedMonth) And Not IsNull(BirthMonth) Then
            If rs!PassedMonth < BirthMonth Then
              Cancel = True
```

Fig. 11. The VBA code added to the *Form_BeforeUpdate* procedure of class *PERSONS* to enforce constraint $C_{20}$

```
         ElseIf rs!PassedMonth = BirthMonth Then
          If Not IsNull(rs!PassedDay) And Not IsNull(BirthDay) Then If rs!PassedDay < BirthDay Then _
            Cancel = True
         End If
        End If
       End If
     End If
    If Not Cancel Then
      rs.MoveNext
    Else
      Beep
      MsgBox Trim(Me!Name) & " was married to " & rs!Name & ", who died in " & rs!PassedYear & " before " & _
            IIf(Sex = "M", "him", "her") & " birth: please change birth date.", vbCritical, "In marriage, both " & _
            "spouses must be simultaneously alive for at least one day!"
    End If
   Wend
   rs.Close
   Set rs = Nothing
  End If
End If
' passed away date > spouse birth date?
If Not Cancel And Not NewRecord And Sex <> "N" And Not IsNull(PassedYear) Then
  If Not IsNull(DLookup("[#M]", "MARRIAGES", "Husband =" & [#R] & " Or Wife =" & [#R])) Then
   Set rs = New ADODB.Recordset            ' current person had at least one spouse
   If Sex = "M" Then                       ' compute the set of wives for the current male person
    q = "SELECT Name, BirthYear, BirthMonth, BirthDay FROM MARRIAGES INNER JOIN RULERS " & _
                                           "ON MARRIAGES.Wife = RULERS.[#R] WHERE Husband =" & [#R]
   Else                                    ' compute the set of husbands for the current female person
    q = "SELECT Name, BirthYear, BirthMonth, BirthDay FROM MARRIAGES INNER JOIN RULERS " & _
                                           "ON MARRIAGES.Husband = RULERS.[#R] WHERE Wife =" & [#R]
   End If
   rs.Open q, Application.CurrentProject.Connection, adOpenForwardOnly, adLockReadOnly
   While Not rs.EOF And Not Cancel         ' check birth dates spouse by spouse
    If Not IsNull(rs!BirthYear) Then       ' null values may not violate this constraint
     If rs!BirthYear > PassedYear Then     ' spouse born after current person's death?
      Cancel = True
     ElseIf rs!BirthYear = PassedYear Then
      If Not IsNull(rs!BirthMonth) And Not IsNull(PassedMonth) Then
       If rs!BirthMonth > PassedMonth Then
        Cancel = True
       ElseIf rs!BirthMonth = PassedMonth Then
        If Not IsNull(rs!BirthDay) And Not IsNull(PassedDay) Then If rs!BirthDay > PassedDay Then _
          Cancel = True
       End If
      End If
     End If
    End If
    If Not Cancel Then
     rs.MoveNext
    Else
     Beep
     MsgBox Me!Name & " was married to " & rs!Name & ", who was born in " & rs!BirthYear & _
       " after " & IIf(Sex = "M", "his", "her") & " death: please change passed away date.", _
       vbCritical, "In marriage, both spouses must be simultaneously alive for at least one day!"
```

Fig. 11. (Continued)

```
    End If
   Wend
   rs.Close
   Set rs = Nothing
  End If
End If
```

Fig. 11. (Continued)

ADODB recordsets were needed in Fig. 11 also because both birth and passed away years, months, and days of persons, as well as marriage and divorce years might be null.

**3.2.2 Code added to the *Form_Current* methods of classes *MARRIAGES* and *WMARRIAGES***

From Fig. 12 it should be noted that this constraint may not be totally enforced preventively in class *MARRIAGES*, for the same reasons as above. However, it is worth eliminating from the *Wife*'s combo-box, to begin with, at least all persons that do not have sex "F" and who were born after the corresponding divorce or husband's passed away years or were dead before the corresponding marriage or husband's birth years. Dually, this constraint may not be totally enforced preventively in the dual class *W_MARRIAGES* either, for the same reasons as above.

However, it is worth eliminating from the *Husband*'s combo-box, to begin with, at least all persons that do not have sex "M" and who were born after the corresponding divorce or wife's passed away years or were dead before the corresponding marriage or wife's birth years. The corresponding code is almost identical: you should only replace *Wife* by *Husband* and 'F' by 'M'.

**3.2.3 Code added to the *Form_BeforeUpdate* method of class *MARRIAGES***

To fully enforce $C_{20}$, the code from Fig. 13 and 14 is absolutely needed in classes *MARRIAGES* (managing the set of wives of the current male person from *PERSONS*) and *W_MARRIAGES* (managing the set of husbands of the current female person from *PERSONS*), respectively.

**3.2.4 Code added to the *Form_BeforeUpdate* method of class *W_MARRIAGES***

Fig. 14 shows the code of the *Form_BeforeUpdate* method of class *W_MARRIAGES*.

# 4 Conclusion and further work

In the framework of the Database Constraint-Driven Design and Development methodology [1], we designed, used, and proposed an algorithm for assisting the process of designing the frontend MS VBA classes of the Windows forms, aimed to enforce both non-relational and relational constraints that are not enforceable by backend relational DBMSes. Where needed and possible, we also provided actual corresponding VBA code patterns. We exemplified the result of applying this algorithm with the VBA code enforcing two such constraints taken from [2].

The algorithm proposed in Section 2, which is the main contribution of this paper, can be easily adapted for similar programming platforms. For example, in MS .Net there exist similar to VBA event-driven procedures, only having different names (e.g., *BeforeUpdate* is called *Validating*, *AfterUpdate* is called *Validated*, etc.).

Dually, thanks to the *de facto* standardization of both ADOX and ODBC, this algorithm may be used without modifications for the design of VBA software frontends based on other relational DBMSes (e.g., Oracle's DB and MySQL, IBM's DB2, Postgres DB, MS Access, etc.): only minor changes in VBA coding are needed (e.g., correspondingly replacing constants storing DBMS error codes, taking care of particular behaviors, like the fact that autonumbering values are generated by MS Access immediately after users type a character on a new data line, which makes them available as early as the *Form_AfterInsert* event-driven procedure, while MS SQL Server does it only after saving the new data line, which makes them available only in the *Form_AfterUpdate* event procedure, etc.).

```
' partially enforce C20 (only for years)
Dim s, w, o As String
s = "SELECT RULERS.[#R], RULERS.Name & IIf(Not IsNull(FATHERS.Name),' of ' &  FATHERS.Name" & _
       ", IIf(Not IsNull(MOTHERS.Name), ' of ' & MOTHERS.Name,)) AS Ruler " _
     & "FROM (RULERS LEFT JOIN RULERS AS FATHERS ON RULERS.Father = FATHERS.[#R]) LEFT " _
       & "JOIN RULERS AS MOTHERS ON RULERS.Mother = MOTHERS.[#R] WHERE RULERS.Sex= 'F' "
o = " ORDER BY RULERS.Name, FATHERS.Name, MOTHERS.Name"
If Not IsNull(MarriageYear) Then
  If Not IsNull(DivorceYear) Then
    w = "And (RULERS.PassedYear Is Null or RULERS.PassedYear >= " & DivorceYear & _
                              ") AND (RULERS.BirthYear Is Null or RULERS.BirthYear <= " & MarriageYear & ")"
  Else
    w = "And (RULERS.PassedYear Is Null or RULERS.PassedYear >= " & MarriageYear & _
                              ") AND (RULERS.BirthYear Is Null or RULERS.BirthYear <= " & MarriageYear & ")"
  End If
Else                          ‘ current marriage year not known
  If Not IsNull(DivorceYear) Then
    w = "And (RULERS.PassedYear Is Null or RULERS.PassedYear >= " & DivorceYear & _
                              ") AND (RULERS.BirthYear Is Null or RULERS.BirthYear <= " & DivorceYear & ")"
  Else
    If Not IsNull(Parent!PassedYear) Then
      If Not IsNull(Parent!BirthYear) Then
        w = "And (RULERS.PassedYear Is Null or RULERS.PassedYear >= " & Parent!BirthYear & _
                      ") AND (RULERS.BirthYear Is Null or RULERS.BirthYear <= " & Parent!PassedYear & ")"
      Else                    ' neither current person's marriage, divorce, or birth years known
        w = "And RULERS.BirthYear Is Null or RULERS.BirthYear <= " & Parent!PassedYear
      End If
    Else                      ' neither current person's marriage, divorce or passed away years known
      If Not IsNull(Parent!BirthYear) Then
        w = "And (RULERS.PassedYear Is Null or RULERS.PassedYear >= " & Parent!BirthYear
      Else                    ' neither current person's marriage, divorce, birth, or passed away years known
        w = ""
      End If
    End If
  End If
End If
Wife.RowSource = s & w & o
Wife.Requery
```

Fig. 12. The VBA code added to the *Form_Current* procedure of class *MARRIAGES* to partially enforce constraint $C_{20}$

```
' enforce C20
Dim v As Variant
Dim wby, wbm, wbd As Integer       ' wife's birth year, month, and day
Dim wpy, wpm, wpd As Integer       ' wife's passed away year, month, and day
Dim wb, wp As Date                 ' wife's birth and passed away dates
Dim hbm, hbd As Integer            ' husband's birth month and day
Dim hpm, hpd As Integer            ' husband's passed away month and day
Dim hb, hp As Date                 ' husband's birth and passed away dates
 If Not Cancel And (NewRecord Or Wife <> Wife.OldValue Or DivorceYear <> DivorceYear.OldValue) Then
  ' C20.1 wife born after divorce or husband's death?
  v = DLookup("BirthYear", "RULERS", "[#R] =" & Wife)
  If Not IsNull(v) Then
    wby = CLng(v)
    If Not IsNull(DivorceYear) Then
      If wby > DivorceYear Then    ' wife born after divorce?
        Cancel = True
        Beep
```

Fig. 13. The VBA code added to the *Form_BeforeUpdate* procedure of class *MARRIAGES* to fully enforce constraint $C_{20}$

```
      MsgBox "For any marriage both spouses must be simultaneously alive for at least one day." & Chr(13) _
           & Wife.Column(1) & " was born in " & wby & " after divorcing!" & Chr(13) & "Please either " & _
           "change divorce year or choose another wife or first change her birth year accordingly.", _
           vbCritical, "Request rejected..."
        DivorceYear.SetFocus
      End If
    Else                              ' divorce year null: wife born after husband's death?
      If Not IsNull(Parent!PassedYear) Then
        v = DLookup("BirthMonth", "RULERS", "[#R] =" & Wife)
        wbm = IIf(IsNull(v), 6, v)        ' when not known wife birth month, assume it was June
        v = DLookup("BirthDay", "RULERS", "[#R] =" & Wife)
        wbd = IIf(IsNull(v), 28, v)       ' when not known wife birth day, assume it was 28 of wbm
        wb = CDate(wby & "-" & wbm & "-" & wbd)
        ' when not known husband death month, assume it was July
        hpm = IIf(IsNull(Parent!PassedMonth), 7, Parent!PassedMonth)
        ' when not known husband death day, assume it was 1st of hpm
        hpd = IIf(IsNull(Parent!PassedDay), 1, Parent!PassedDay)
        hp = CDate(Parent!PassedYear & "-" & hpm & "-" & hpd)
        If DateDiff("d", wb, hp) < 0 Then
          Cancel = True
          Beep
          MsgBox "For any marriage, both spouses must be simultaneously alive for at least one day." _
            & Chr(13) & Parent!Name & " passed away on " & hp & ", while " & Wife.Column(1) _
            & " was born after that, on " & wb & "!" & Chr(13) & "Please choose another wife or " _
            & "first change these spouses passed away or/and birth dates accordingly.", vbCritical, _
            "Request rejected..."
          Wife.SetFocus
        End If
      End If
    End If
  End If
End If
If Not Cancel And (NewRecord Or Wife <> Wife.OldValue Or MarriageYear <> MarriageYear.OldValue) Then
  ' C20.2 wife dead before marriage date or husband's birth?
  If Not Cancel Then
    v = DLookup("PassedYear", "RULERS", "[#R] =" & Wife)
    If Not IsNull(v) Then
      wpy = CLng(v)
      If Not IsNull(MarriageYear) Then
        If wpy < MarriageYear Then     ' wife passed away before marriage?
          Cancel = True
          Beep
          MsgBox "For any marriage, both spouses must be simultaneously alive for at least one day." _
              & Chr(13) & Wife.Column(1) & " passed away in " & wpy & " before marrying!" & Chr(13) & _
              "Please either change marriage year or choose another wife or first change her passed away " & _
              "year accordingly.", vbCritical, "Request rejected..."
          MarriageYear.SetFocus
        End If
      Else                            ' marriage year null: wife passed away before husband's birth?
        If Not IsNull(Parent!BirthYear) Then
          v = DLookup("PassedMonth", "RULERS", "[#R] =" & Wife)
          wpm = IIf(IsNull(v), 7, v)      ' when not known wife passed away month, assume it was July
          v = DLookup("PassedDay", "RULERS", "[#R] =" & Wife)
          wpd = IIf(IsNull(v), 1, v)      ' when not known wife passed away day, assume it was 1st of wpm
          wp = CDate(wpy & "-" & wpm & "-" & wpd)
          ' when not known husband birth month, assume it was June
          hbm = IIf(IsNull(Parent!BirthMonth), 6, Parent!BirthMonth)
          ' when not known husband birth day, assume it was 28 of hbm
          hbd = IIf(IsNull(Parent!BirthDay), 28, Parent!BirthDay)
          hb = CDate(Parent!BirthYear & "-" & hbm & "-" & hbd)
```

Fig. 13. (Continued)

```
        If DateDiff("d", hb, wp) < 0 Then
          Cancel = True
          Beep
          MsgBox "For any marriage, both spouses must be simultaneously alive for at least one day." _
            & Chr(13) & Parent!Name & " was born on " & hb & ", while " & Wife.Column(1) _
            & " passed away before that, on " & wp & "!" & Chr(13) & "Please choose another wife or " _
            & "first change these spouses passed away or/and birth dates accordingly.", vbCritical, _
            "Request rejected..."
          Wife.SetFocus
        End If
      End If
    End If
  End If
 End If
End If
```

Fig. 13. (Continued)

```
' enforce C20
Dim v As Variant
Dim wby, wbm, wbd As Integer  ' wife's birth year, month, and day
Dim wpy, wpm, wpd As Integer  ' wife's passed away year, month, and day
Dim wb, wp As Date            ' wife's birth and passed away dates
Dim hbm, hbd As Integer       ' husband's birth month and day
Dim hpm, hpd As Integer       ' husband's passed away month and day
Dim hb, hp As Date            ' husband's birth and passed away dates
If Not Cancel And (NewRecord Or IsNull(Husband.OldValue) Or Husband <> Husband.OldValue Or
    IsNull(DivorceYear.OldValue) Or DivorceYear <> DivorceYear.OldValue) Then
  ' C20.1 Husband born after divorce or wife's death?
  v = DLookup("BirthYear", "RULERS", "[#R] =" & Husband)
  If Not IsNull(v) Then
    hby = CLng(v)
    If Not IsNull(DivorceYear) Then
      If hby > DivorceYear Then        ' Husband born after divorce?
        Cancel = True
        Beep
        MsgBox "For any marriage, both spouses must be simultaneously alive for at least one day." & _
          Chr(13) & Husband.Column(1) & " was born in " & wby & " after divorcing!" & Chr(13) & _
          "Please either change divorce year or choose another husband or first change his birth year " & _
          "accordingly.", vbCritical, "Request rejected..."
        DivorceYear.SetFocus
      End If
    Else                               ' divorce year null: husband born after wife's death?
      If Not IsNull(Parent!PassedYear) Then
        v = DLookup("BirthMonth", "RULERS", "[#R] =" & Husband)
        hbm = IIf(IsNull(v), 6, v)     ' when not known husband birth month, assume it was June
        v = DLookup("BirthDay", "RULERS", "[#R] =" & Husband)
        hbd = IIf(IsNull(v), 28, v)    ' when not known husband birth day, assume it was 28 of hbm
        hb = CDate(hby & "-" & hbm & "-" & hbd)
        ' when not known wife death month, assume it was July
        wpm = IIf(IsNull(Parent!PassedMonth), 7, Parent!PassedMonth)
        ' when not known wife death day, assume it was 1st of wpm
        wpd = IIf(IsNull(Parent!PassedDay), 1, Parent!PassedDay)
        wp = CDate(Parent!PassedYear & "-" & hpm & "-" & hpd)
        If DateDiff("d", hb, wp) < 0 Then
          Cancel = True
          Beep
          MsgBox "For any marriage, both spouses must be simultaneously alive for at least one day." _
            & Chr(13) & Parent!Name & " passed away on " & wp & ", while " & Husband.Column(1) _
            & " was born after that, on " & hb & "!" & Chr(13) & "Please choose another husband or " _
            & "first change these spouses passed away or/and birth dates accordingly.", vbCritical, _
            "Request rejected..."
```

Fig. 14. The VBA code added to the *Form_BeforeUpdate* procedure of dual class *W_MARRIAGES* to fully enforce constraint $C_{20}$

```
          Husband.SetFocus
        End If
      End If
   End If
  End If
End If
If Not Cancel And (NewRecord Or Husband <> Husband.OldValue Or IsNull(MarriageYear.OldValue) Or
        MarriageYear <> MarriageYear.OldValue) Then
   'C20.2 husband dead before marriage date or wife's birth?
  If Not Cancel Then
    v = DLookup("PassedYear", "RULERS", "[#R] =" & Husband)
    If Not IsNull(v) Then
      hpy = CLng(v)
      If Not IsNull(MarriageYear) Then
        If hpy < MarriageYear Then      ' husband passed away before marriage?
          Cancel = True
          Beep
          MsgBox "For any marriage, both spouses must be simultaneously alive for at least one day." & _
            Chr(13) & Husband.Column(1) & " passed away in " & hpy & " before marrying!" & Chr(13) & _
            "Please either change marriage year or choose another husband or first change his passed " & _
            "away year accordingly.", vbCritical, "Request rejected..."
          MarriageYear.SetFocus
        End If
      Else                                ' marriage year null: husband passed away before wife's birth?
        If Not IsNull(Parent!BirthYear) Then
          v = DLookup("PassedMonth", "RULERS", "[#R] =" & Husband)
          hpm = IIf(IsNull(v), 7, v)        ' when not known husband passed away month, assume it was July
          v = DLookup("PassedDay", "RULERS", "[#R] =" & Husband)
          hpd = IIf(IsNull(v), 1, v)        ' when not known husband passed away day, assume it was 1st of hpm
          hp = CDate(hpy & "-" & hpm & "-" & hpd)
         ' when not known wife birth month, assume it was June
          wbm = IIf(IsNull(Parent!BirthMonth), 6, Parent!BirthMonth)
          ' when not known wife birth day, assume it was 28 of wbm
          wbd = IIf(IsNull(Parent!BirthDay), 28, Parent!BirthDay)
          wb = CDate(Parent!BirthYear & "-" & hbm & "-" & hbd)
          If DateDiff("d", wb, hp) < 0 Then
            Cancel = True
            Beep
            MsgBox "For any marriage, both spouses must be simultaneously alive for at least one day." _
              & Chr(13) & Parent!Name & " was born on " & wb & ", while " & Husband.Column(1) _
              & " passed away before that, on " & hp & "!" & Chr(13) & "Please choose another husband or " _
              & "first change these spouses passed away or/and birth dates accordingly.", vbCritical, _
              "Request rejected..."        Husband.SetFocus
          End If
        End If
      End If
    End If
  End If
End If
```

Fig. 14. (Continued)

Moreover, our proposed algorithm could also be used for software applications built on top of NoSQL DBMSes: as they lack support for even almost all the relational constraint types, the only difference is that the set of the relational constraints to be enforced by the frontend is much larger, but the algorithm of enforcing them is the same as for the relational DBMS ones.

In [2] we also analyzed 8 of the top 10 genealogical software applications of 2023, as considered by [15] (the other 2 are unavailable to us: one is meant only for Apple platforms and the other has no free version). Unfortunately, only a few of them are at most issuing a couple of warnings when trying to save unplausible data, which you can dismiss and save it, however. All of them have glossy interfaces, but are ok with managing aberrant data like people being born after their christening, last marriage, having children, and even being dead, or people being buried before death, or living some negative years, or huge positive ones, etc.

In sharp contrast, the *GENEALOGIES* application proposed in [2], which uses the DBCDDD methodology, and the algorithm proposed in this paper, provides the highest possible quality of its managed data, which is satisfying all constraints governing this sub-universe.

Further work could be done: first, to exemplify the power of our proposed algorithm even in the frameworks using NoSQL DBMSes (e.g., MariaDB, MongoDB, Cassandra, Oracle NoSQL DB, etc.); secondly, devising similar algorithms for other SQL embedding platforms (e.g., C#, Java, Python, etc.).

**Acknowledgement:** This work was supervised by Associate Professor Cristina Șerban, from the Math. & Computer Science Dept. of the Ovidius University at Constanta, Romania.

DIANA CHRISTINA MANCAS
Ovidius University
Mathematics and Computer Science Department
Bd. Mamaia 123, Constanta
ROMANIA
diana.mancas@365.univ-ovidius.ro